# Hollow-Core Terahertz Optical Waveguides With Hyperuniform Disordered Reflectors

Tian Ma[1], Martin Girard[1,2], Andrew Squires[3], Roger Lewis[3], Roorda Sjoerd[4] and Maksim Skorobogatiy[1*]

[1]*Department of Engineering Physics, École Polytechnique de Montréal, Montreal, Québec, H3T 1J4, Canada*
[2]*2Department of Materials Science and Engineering, Northwest University, Evanston, IL 60208, USA*
[3]*Institute for Superconducting and Electronic Materials, School of Physics, University of Wollongong, NSW 2522, Australia*
[4]*Department of Physics, Université de Montréal, Montreal, Québec, H3T 1J4, Canada*

*\*Maksim.skorobogatiy@polymtl.ca*

**Abstract:** Novel hollow-core THz waveguides featuring hyperuniform disordered reflectors are proposed, fabricated, and characterized. The reflector comprise aperiodically positioned dielectric cylinders connected with dielectric bridges. The proposed waveguides are fabricated using a 3D stereolithography printer. Optical properties of the fabricated waveguide are investigated numerically using finite element method, as well as experimentally using THz-TDS spectroscopy. The results confirm that proposed waveguides exhibit sizable photonic band gaps (20%) even when relatively low refractive index contrast used (resin/air). Position of the band gaps can be easily tuned by varying reflector geometrical parameters.

## 1. Introduction

Photonics crystal (PC) materials have drawn great interest over the years because of their unique properties that allow advanced light management [1]. In particular, dielectric reflectors based on PCs can be employed to create hollow-core fibers by arranging such reflectors around a gas filled cavity. In such fibers, the light is confined in the hollow-core for frequencies within the reflector photonic band gaps (PBGs). Based on this principle, various hollow-core PBG fibers have been proposed for simultaneously low-loss and low-dispersion guidance over sizable spectral ranges [2-3]. These fibers can be divided into two categories, Bragg fibers and holey fibers, according to their reflector structure.

Generally, hollow core Bragg fibers consist of a circularly symmetric Bragg reflector which is formed by alternate high and low refractive index layers. The Bragg reflector can be all-solid or porous. The all-solid Bragg reflector is formed by repeating alternate concentric layers of two different polymers [4] or the same polymer with different dopants [5-6]. The bandgap position and its spectral width are determined by the thickness and the refractive indices of the alternate layers. The periodic refractive index variation in Bragg reflectors can also be realised by introducing rings of porous material. Using this approach, hollow core Bragg fibers with solid/randomly porous multilayers [7], air-hole rings [8] and cob-web structures [9-10] were investigated in the terahertz range. Another type of the hollow core PBG fibers is a holey fiber. Such fibers feature reflectors formed by various types of periodic lattices, such as, rectangular [11], triangular [11, 12], honeycomb [13], etc. The holey PBG fibers are typically designed to have high air-filling fractions in order to achieve bandgap.

Recently, both numerical [14-16] and experimental [17-20] studies in 2D have shown that hyperuniform disordered structures present a new class of disordered photonics materials that can possess large complete photonic band gaps for all polarizations. In these studies, the key parameter that characterized hyperuniform structures is the hyperuniformity χ, which was first introduced as an order metric of a point pattern based on its local density fluctuations [19]. The hyperuniformity ranges from 0 (disordered) to 1 (crystalline). A particular type of a hyperuniform disordered structure that was considered in [19] comprises dielectric cylinders connected by the thin dielectric bridges. Based on this structure, various planar hyperuniform waveguides have been developed with both high [16] and low refractive index contrasts [17] that exhibited spectrally broad bandgaps, as well as photonic bandgap guidance for all polarizations. Moreover, it was demonstrated in [18] that for the same refractive index contrast, hyperuniform reflectors can have larger bandgaps than their counterparts featuring periodic PCs. Thus, it could be expected that hollow core PBG fibers featuring hyperuniform reflectors could have spectrally broader bandgap than hollow core PBG fibers that use strictly periodic reflectors.

In this paper, we propose a novel hollow core terahertz PBG waveguide that uses hyperuniform disordered reflectors. This is essentially a generalization of the earlier 2D waveguides featuring hyperuniform claddings [14-20] into 3D waveguides and fibers. Our main motivation is to explore the possibility of designing hollow core waveguides that feature spectrally broad bandgaps that are potentially superior to those attainable with purely periodic structures. Particularly, we demonstrated theoretically that using resin/air material combination that offers relatively low refractive index contrast of 1.67/1, one can design a hollow core waveguide featuring an 80GHz (~20%) bandgap centered in the vicinity of 0.4THz. In such waveguide, highly porous PBG reflector comprised ~113μm radius cylinders connected with ~35μm thick bridges. We then attempted fabrication of such waveguides using 3D stereolithography. The diameter of the resultant waveguides (reflector size) is ~20mm, while the diameter of the hollow core is ~5mm. Due to limitations of 3D printer used in our work, the resolution was limited to 100μm which allowed us to print structures with bridges thicker than 200μm. As we demonstrated both theoretically and experimentally, thicker bridges lead to the overall reduction in the bandgap spectral size. Nevertheless, the fabricated waveguides featured relative wide bandgaps (up to ~15%), and low transmission losses (<0.10cm$^{-1}$) within their PBGs.

## 2. Fiber design and fabrication

In order to generate a disordered reflector structure, we use a set of dielectric cylinders connected with thin dielectric bridges. The cylinder centers follow a 2D *hyperuniform* point pattern. For any point pattern, its point distribution can be characterized by its number variance, which is given by the standard deviation of the number of points ($N_R$) in a sampling window Ω of radius $R$ in $d$ dimension, $\sigma^2(R) = \langle N_R^2 \rangle - \langle N_R \rangle^2$ [21]. As an example, in 3D sampling window Ω is a sphere, while in 2D it is a circle. A point pattern is called "*hyperuniform*" if the corresponding number variance within Ω grows more slowly than the volume of Ω namely $R^d$. For example, if a point distribution pattern in 2D has the number variance $\sigma^2(R) \sim R$, then such a pattern is hyperuniform. In reciprocal space, the point distribution can be characterized by its structure factor $S(\boldsymbol{k})$, which can be given as

$$S(\boldsymbol{k}) = \frac{1}{N} \left| \sum_{j=1}^{N} \exp[-i\boldsymbol{k} \cdot \boldsymbol{r}_j] \right|^2 = \frac{|\rho(\boldsymbol{k})|^2}{N} \tag{1}$$

where $r_j$ refers to the location of particle $j$, while $\rho(\mathbf{k})$ is known as the point pattern collective coordinates defined as the Fourier transfer of the particle locations. For a hyperuniform pattern, S($\mathbf{k}$) tends to 0 when the wavenumber $|\mathbf{k}|$ approaches 0. Moreover, the point pattern is called "stealthy" when the structure factor is isotropic, continuous and equal to 0 in the range of $|\mathbf{k}| < k_C$ for some positive $k_C$. For the resultant hyperuniform pattern, its hyperuniformity parameter, χ, is defined as a ratio of the number of constrained degrees of freedom $M(k_C)$ to the total number of degrees of freedom $d \cdot N$ where $N$ is the total number of points and $d$ is the number of dimensions [20]

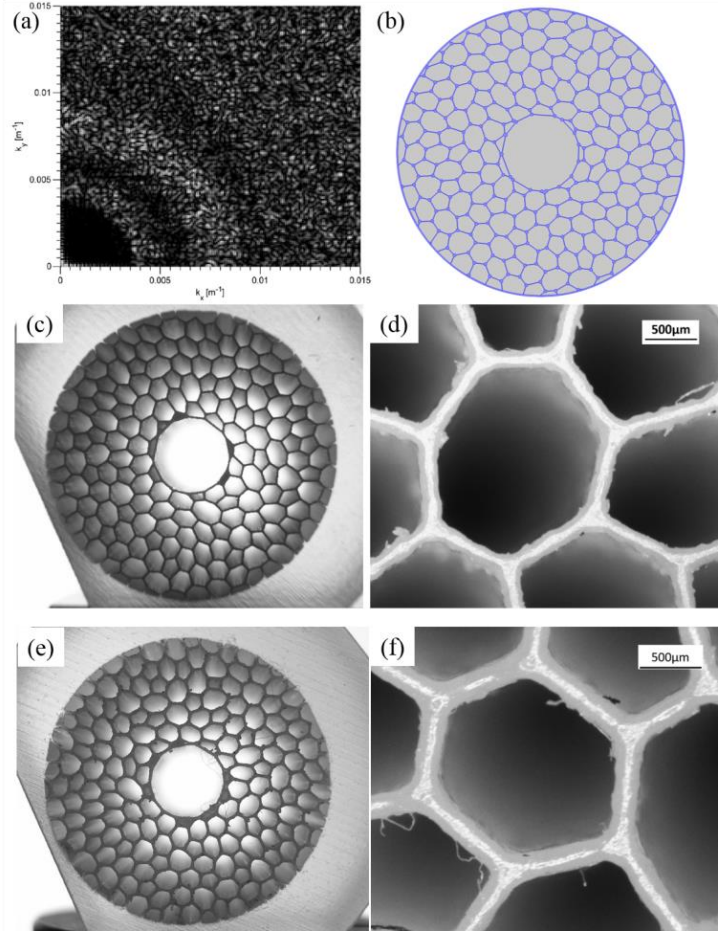

Fig. 1. (a) Hyperuniform point pattern in k-space. This point pattern is used to define center positions of the dielectric cylinders in the hyperuniform PBG reflector. (b) Waveguide and a simulation cell used in our numerical calculation. The reflector material is shown in blue, while the air is gray and it is terminated by a circular perfect electric conductor (PEC). The cylinder radii are 113µm and the bridges thicknesses 35µm. (c) The fabricated waveguide with the bridge thickness of 200µm in the cladding. (d) Zoom of the reflector region shown in (c). (e) The fabricated waveguide with the bridge thickness of 250µm. (f) Zoom of the reflector region shown in (e).

$$\chi = \frac{M(k_C)}{d \cdot N} \quad (2)$$

As detailed in [14], with $\chi = 0.5$, the proposed hyperuniform structure can be optimized to exhibit complete PBG for both TE and TM polarizations. In our design, we used the same hyperuniformity parameter $\chi = 0.5$ as in [14], and the point pattern was generated from a randomly distributed point pattern (uniform distribution) consisting of $N$ points in a square box with the side length of $L$ by minimizing its total potential energy, which is defined using the pairwise potential $v(\boldsymbol{r_j} - \boldsymbol{r_i})$ that describes the interaction between particles $i$ and $j$

$$\Phi = \sum_{i=1}^{N}\sum_{j<i} v(r_j - r_i)$$

$$= \Omega^{-1} \sum_{\boldsymbol{k}} V(\boldsymbol{k}) \sum_{i=1}^{N}\sum_{j<i} \cos[\boldsymbol{k} \cdot (\boldsymbol{r_j} - \boldsymbol{r_i})]$$

$$= \Omega^{-1} \sum_{\boldsymbol{k}} V(\boldsymbol{k}) C(\boldsymbol{k}) \quad (3)$$

In the expression above, $V(\boldsymbol{k})$ is the Fourier transform of the pairwise potential,

$$V(\boldsymbol{k}) = \int d\boldsymbol{r}\, v(\boldsymbol{r})\exp(i\boldsymbol{k}\boldsymbol{r}) \quad (4)$$

While $C(\boldsymbol{k})$ is a real function that can be written using the structure factor $S(\boldsymbol{k})$ as follow

$$C(\boldsymbol{k}) = \sum_{i=1}^{N}\sum_{j<i} \cos[\boldsymbol{k} \cdot (\boldsymbol{r_i} - \boldsymbol{r_j})]$$

$$= \frac{1}{2}[\rho(\boldsymbol{k})\rho(-\boldsymbol{k}) - \rho(0)]$$

$$= \frac{N}{2}(S(\boldsymbol{k}) - 1) \quad (5)$$

While various forms of the pairwise interaction potential are possible, for simplicity, we use the 'square mound' potential in the k-space similarly to [14, 20-21]

$$V(\boldsymbol{k}) = \begin{cases} V_0 > 0, & \text{for } |\boldsymbol{k}| \leq \boldsymbol{k_c} \\ 0, & \text{otherwise} \end{cases} \quad (6)$$

In the infinite-volume limit, the corresponding real space pairwise potential is given by [21]

$$\lim_{\Omega \to \infty} v(\boldsymbol{r}) = (V_0 k_c / 2\pi r) J_1(k_c r) \quad (7)$$

where $J_1$ is the first order of the Bessel function of the first kind. With this choice of the pairwise potential, the total potential energy $\Phi$ can be minimized by driving $C(\boldsymbol{k})$ or $S(\boldsymbol{k})$ to its minimum value for all $|\boldsymbol{k}| \leq k_c$. The minimization is realised by changing the particle coordinates using the TOMLAB's MINOP algorithm, which is a Fortan-based reduced gradient nonlinear optimization solver. In Fig. 1(a), we illustrate the Fourier transform in the k space of the generated hyperuniform point pattern. The brightness of each point is proportional to the value of its structure factor $S(\boldsymbol{k})$.

Thus, following the method described in [14], we developed the cross section of the proposed waveguide based on the generated hyperunifrom point pattern. The triangular mesh is defined with the hyperuniform point pattern as its vertices using the Delaunay triangulation method. Then, cylinders with the radii of $r_c$ are placed at the centroid of each triangular cell. Finally, the cylinders in neighboring triangular cells are connected using dielectric bridges of thickness $t_b$. The central part of thus generated structure was replaced with a hollow core of 5mm diameter. The final step in our design was to maximize the full PBG width of the proposed waveguide by optimizing its structural parameters, namely the cylinder radius $r_c$ and the bridge thickness $t_b$. Similar optimization has been done in [14], where these structural parameters were optimized for a planar hyperuniform waveguide with the optimized parameters expressed as

$$r_c = \alpha \frac{L}{\sqrt{N}}$$
$$t_b = \beta \frac{L}{\sqrt{N}} \tag{8}$$

where *L* is the size length of the supercell and N is the number of points in this supercell. For our waveguide, *L=21mm* and *N=256*. In our simulation, we set the central frequency of the PBG at 0.4THz and the cladding material refractive index at 1.67. Then, by performing consecutive sweeps of both $\alpha$ and $\beta$ parameters, we can iteratively optimize the waveguide structure and maximise the resultant PBG at a fixed frequency. Particularly, at each optimization step we fix one of the parameters (say α) at the optimal value found in the previous step. Then, we perform a one dimensional sweep of the other parameter (β) and find its new optimal value. We then repeat the procedure by switching the parameters (fix β, sweep α). Optimal value of a parameter is defined as one that results in the equidistant separation of the air light line from both the lower and the upper edges of the continuum of cladding states. This optimisation condition is meant to minimize scattering of the core guided modes (with effective refractive indices close to that of air) into the continuum of cladding modes. After several such iterations the values of the two parameters converge to their optimal values of $\alpha = 0.084$ and $\beta = 0.027$ with the corresponding optimal cylinder radius and bridge thickness being 113µm and 35µm, respectively. In Fig. 2, we demonstrate two consecutive sweeps of $\alpha$ and $\beta$ after convergence is achieved. The proposed waveguide with optimized parameters is shown in Fig. 1(b), and the corresponding band diagram is demonstrated in Fig. 4. The resultant band gap width is ~80GHz.

We then attempted fabrication of these developed waveguides using a commercial stereolithography 3D printer (ProJet® 3500HD Plus) with the photoresin (*VisyJet® Crystal*). To print robust structures, with this printer, the minimum feature size in the lateral directions is required to be at least twice the resolution ~200µm. Hence we could not print the waveguide with the optimized structure. Instead of the optimized structure, we therefore printed 2 types of waveguides with bridge thicknesses of 200µm and 250µm, respectively, while keeping the same distribution of the cylinders. For each waveguide, we printed 4 sections with the length of 25mm each. As we used the same drawing file for the printing, these two waveguides have almost the same reflector size with ~22mm outer diameter and ~5mm diameter hollow core. The cross-section of the fabricated waveguides are demonstrated in Fig. 1(c) and (e).

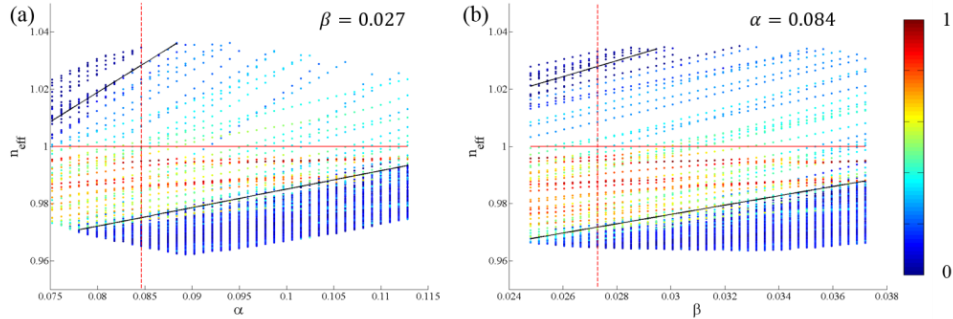

Fig. 2 Optimized waveguide structure. (a) Sweeping $\alpha$ with the fixed $\beta = 0.027$ results in the optimal value $\alpha = 0.084$, while (b) sweeping $\beta$ with the fixed $\alpha = 0.084$ results in the optimal value $\beta = 0.027$. Two black solid lines define boundaries of the continuum of cladding-bound states. The red line refers to the air light line with n=1. The red dashed line shows the optimal parameter value for which the air light line is positioned strictly in the middle between the two boundaries with the continuum of cladding states.

## 3. Optical Characterization

### 3.1 Optical characterization of the reflector material (*VisyJet® Crystal*)

In order to model the waveguide optical properties, firstly, we characterized the refractive index and absorption losses of the reflector material (*VisyJet® Crystal*) used in waveguide fabrication. Characterization of the refractive index and absorption losses were performed using Zomega THz time domain spectrometer (THz-TDS) [22] using thin resin slices of different thicknesses, which were prepared using the same 3D printer. In the experiment, the complex transmission of these resin slices with different thicknesses are measured.

The real part of the resin refractive index and the absorption coefficient were extracted from the measured complex transmission data. In the analysis, multiple reflections within the sample were neglected. In this case, the measured complex transmission is given by [23]:

$$T(\omega, L) = \frac{E_t}{E_r} = |T(\omega, L)| \cdot \exp[i\varphi(\omega, L)]$$

$$|T(\omega, L)| = C_{in} \cdot C_{out} \cdot \exp\left[-\frac{\alpha(\omega)L}{2}\right] \quad (9)$$

$$\varphi(\omega, L) = -\frac{\omega}{c}(n_r(\omega) - 1)L$$

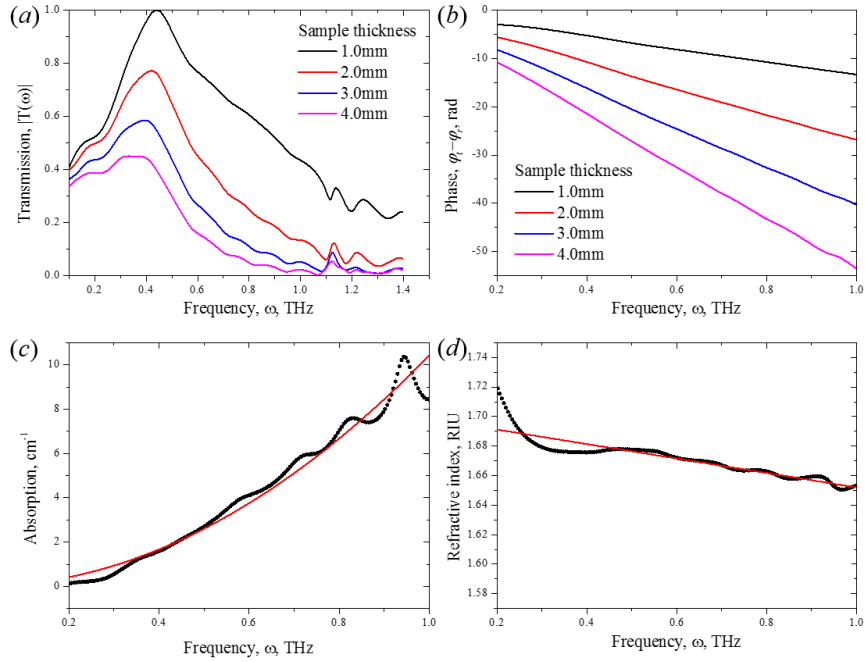

Fig. 3 Optical characterization of the reflector material using cutback method and four 1mm-thick slices of the *VisyJet®* Crystal resin. (a) Normalized transmission spectra, (b) corresponding unwrapped phases, (c) the absorption coefficient and (d) the real part of the reflective index.

where $E_t$ and $E_r$ are complex transmission spectra measured with and without the tested sample with the length of $L$. $C_{in}$ and $C_{out}$ are the input and output coupling coefficient with respect to the tested sample, which are assumed to be the same during all samples. $n_r$ and $\alpha$ are the real part of the reflective index and the absorption coefficient of the tested sample, respectively. In Fig. 3 (a) and (b), we present transmission spectra and corresponding phases of the tested samples with different thicknesses. Our analysis is limited to the frequency range of 0.2-1.0THz, where all the spectra are well above the noise level. For these frequencies, the absorption coefficient extracted from the measured data [see Fig. 2(c)] can be fitted with a second order polynomial with respect to frequency:

$$\alpha(\omega)[cm^{-1}] = 10.43 \cdot (\omega[THz])^2 \qquad (10)$$

In Fig. 2(d), we demonstrate the frequency-dependent refractive index of the reflector material, which is retrieved using the unwrapped phase relative to the reference [see Fig. 2(b)]. In the frequency range of 0.2-1.0THz, the refractive index of the reflector material decreases almost linearly towards higher frequencies and it can be fitted as:

$$n(\omega) = 1.701 - 0.049 \cdot \omega[THz] \qquad (11)$$

3.2 Band diagram of the proposed waveguides

Light guidance in the proposed waveguides was analysed using commercial finite element software COMSOL. For the experimentally fabricated waveguides, the reflector geometries were extracted from the high resolution photographs of the waveguides cross-sections [see Fig. 1 (c) and (e)]. For the frequency dependent refractive index and absorption loss of the reflector material, we used polynomial fits presented by Eq. 7 and 8. Computational cell was terminated by the circular perfect electric conducting boundary. Modal dispersion relations of all guided modes for the two fabricated waveguides with different bridge thicknesses are presented in Fig. 5. In these band diagrams, we present the modal effective refractive indices ($n_{eff}$) of the guided modes as a function of frequency in the range of 0.1THz-0.5THz. Due to large system size and small features, modal simulation above 0.5THz is problematic due to time and memory limitations. The color code in Fig. 4 and 5 indicates the fraction of the power guided by the individual mode within the hollow core. Thus, the blue color refers to modes with the modal localisation mostly outside of the waveguide core, while the red color refers to strong presence of the modal fields in the hollow core. In order to show core guided modes clearly, we use bigger dots to represent the modes with more than 60% of the total power in the core. The red solid line in these diagrams is the light line of air with $n = 1$, while the red dashed lines define the edges of the photonic bandgaps.

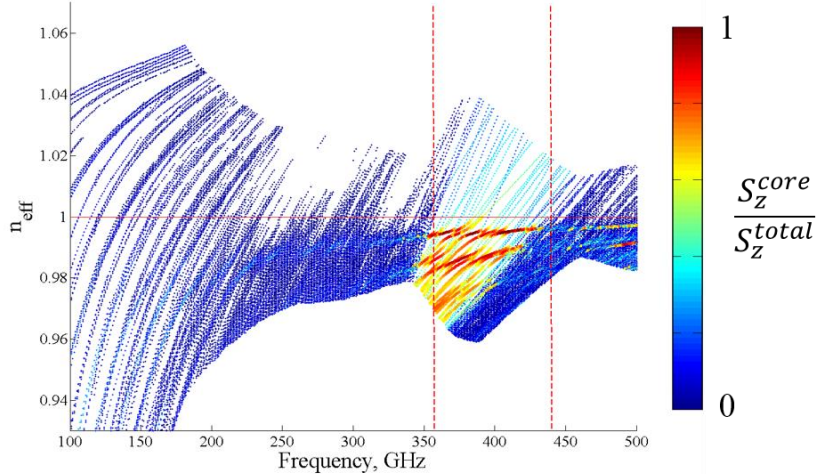

Fig. 4 Band diagram of the waveguide with numerically optimized reflector structure. Color of each dot indicates the fraction of power guided in the hollow core.

As discussed in Section 2, the structure of the proposed waveguide has been optimized to enlarge its PBGs. The waveguide with numerically optimized reflector structure ($r_c = 113\mu m, t_b = 35\mu m$) exhibits a relatively wide photonic band gap of ~80GHz, which is centered at 0.4THz. Within this band gap, the guided modes are strongly confirmed to the waveguide hollow core, while for frequencies outside of the band gap, guided modes show strong presence in the cladding. As fabricated waveguides feature suboptimal bridge and cylinder sizes, resultant band gaps are smaller and positioned at lower frequencies. As seen in Fig. 4, the two fabricated waveguides exhibit several PBGs in the frequency range of 0.1-0.5THz. For the fabricated waveguide with bridge thickness of 200μm, four main band gaps are centered at 0.14THz, 0.24THz, 0.37THz, and 0.45GHz, with the band widths of 13GHz, 12GHz, 40GHz, 14GHz, respectively. At other frequencies, such as 0.32GHz and 0.43GHz, some minor PBGs are also exhibited. As the bridge thickness is increased to 250μm, central frequencies of these PBGs are further shifted to lower frequencies. In this case, some PBGs are discernible at 0.13THz, 0.24THz, 0.33THz and 0.41GHz with corresponding band widths of 20GHz, 14GHz, 25GHz and 30GHz.

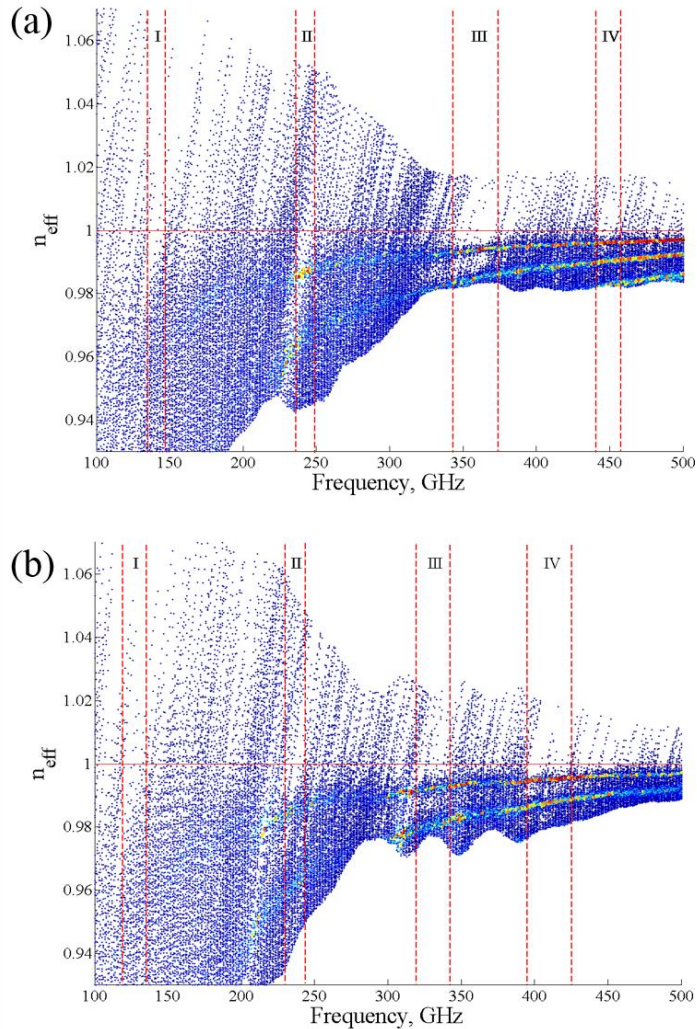

Fig. 5 Band diagrams for the two fabricated waveguides with the bridge thicknesses of (a) 200μm and (b) 250μm. The color scale is the same as the one presented in Fig. 3.

In Fig. 6, we demonstrated examples of the longitudinal flux ($S_z$) distributions for the two modes with frequencies inside of the PBG (0.41THz) and outside of the PBG (0.35THz) (Here, we consider waveguide with numerically optimized geometry shown in Fig. 1(b)). Clearly, the mode with a frequency inside of the reflector band gap is tightly confined to the waveguide core, while the mode with a frequency outside of the reflector bandgap has a strong presence in the cladding.

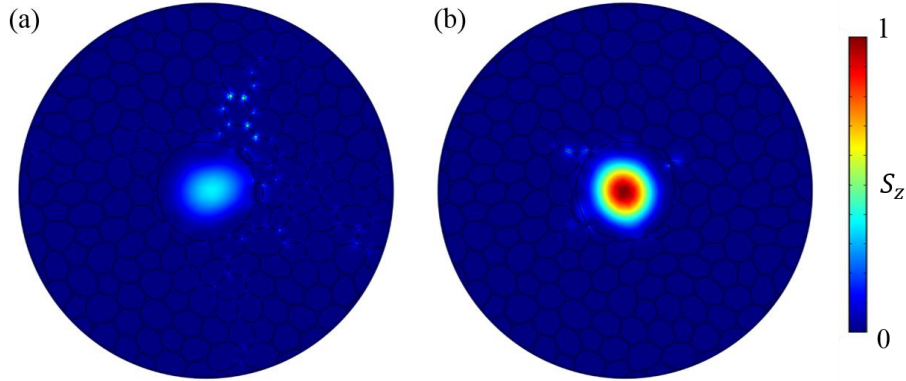

Fig. 6. Examples of the longitudinal flux distributions for the modes of an optimized waveguide (a) outside of the PBG, $f = 0.35$THz, and (b) inside of the PBG, $f = 0.41$THz

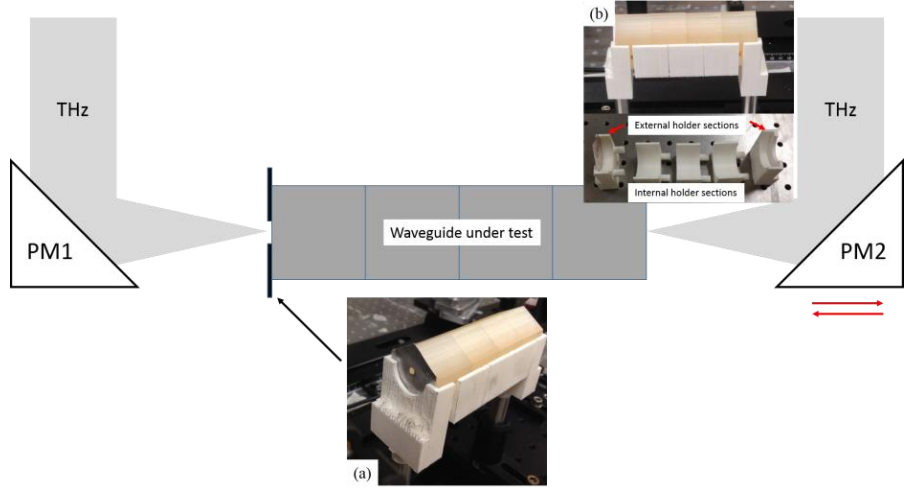

Fig. 7. Schematic of the experimental setup with all of four waveguide sections mounted inside of a composite holder. Insert (a): the input facet of a waveguide features an aperture with the size equal to that of the waveguide hollow core. Insert (b): the 3D printed composite holder with sections before (bottom) and after (top) being insert into the experimental setup. PM1 – fixed parabolic mirror with a focus at the waveguide input edge. PM2 – movable parabolic mirror which is displaced every time the waveguide section is removed in order to keep the focal point at the waveguide output edge.

*3.3 Waveguide transmission measurements*

Next, we characterized THz transmission of the fabricated waveguides using a modified terahertz time-domain spectroscopy (THz-TDS) setup shown in Ref. [23]. The setup consists of a frequency doubled femtosecond fiber laser (MenloSystem® C-fiber laser) used as the pump source and two identical GaAs dipole antenna (MenloSystem®) used as THz emitter and detector operating in the 0.1-3.0THz range. Fig. 7 illustrates the experimental setup where the waveguide under test is fixed in the U-shaped 3D printed composite holder sections placed between the two parabolic mirrors. Here, the printed holder sections can be divided into two types, namely external- and internal-holder sections [see Fig. 7(b)]. Two external holder sections were used to position the input and output facets of the waveguide with respect to the parabolic mirrors. The first external holder section was fixed in the focal point of the stationary parabolic mirror (PM1), thus, ensuring that the input facet of the waveguide is also at the focal point of PM1. The second external holder was mounted together with the parabolic mirror 2 (PM2) on the movable stage. At the same time, the second holder was fixed in the focal point of PM2. The internal holder sections hosting the waveguide were sandwiched between the two external holder sections, thus ensuring that that the waveguide input and output ends were always in the focal points of the two parabolic mirrors. During the measurements, the three internal holder sections that each hosted 2.5cm-long waveguide sections, were removed one by one, and hence, waveguides of 10cm, 7.5cm, 5cm and 2.5 cm were measured. At the input facet, an aluminum foil with a central hole of 5mm diameter was glued on the fixed external holder section, and it was used as an aperture to prevent the incoming beam from coupling into the waveguide cladding.

In Fig. 8, we demonstrate the experimentally measured electric field time domain THz traces and corresponding transmission spectra of the fabricated waveguides. The references in Fig. 8 were acquired by removing all the internal holder sections (together with the waveguide sections), and minimizing the distance between the two external holder sections. The corresponding transmission spectra normalised with respect to the references are shown in Fig. 9. Here, the waveguide I features 200µm thick bridges, [Fig. 1(c) and (d)], while the waveguide II has 250µm thick bridges [Fig. 1(e) and (f)]. In order to make comparison with our theoretical simulations, we focus on the transmission properties of fabricated waveguide in the frequency range of 0.1THz-0.5THz. We note that the band gaps become fully formed only when the length of the tested waveguide is longer than 5.0cm, while when only one section is used, the waveguide rather functions as an aperture.

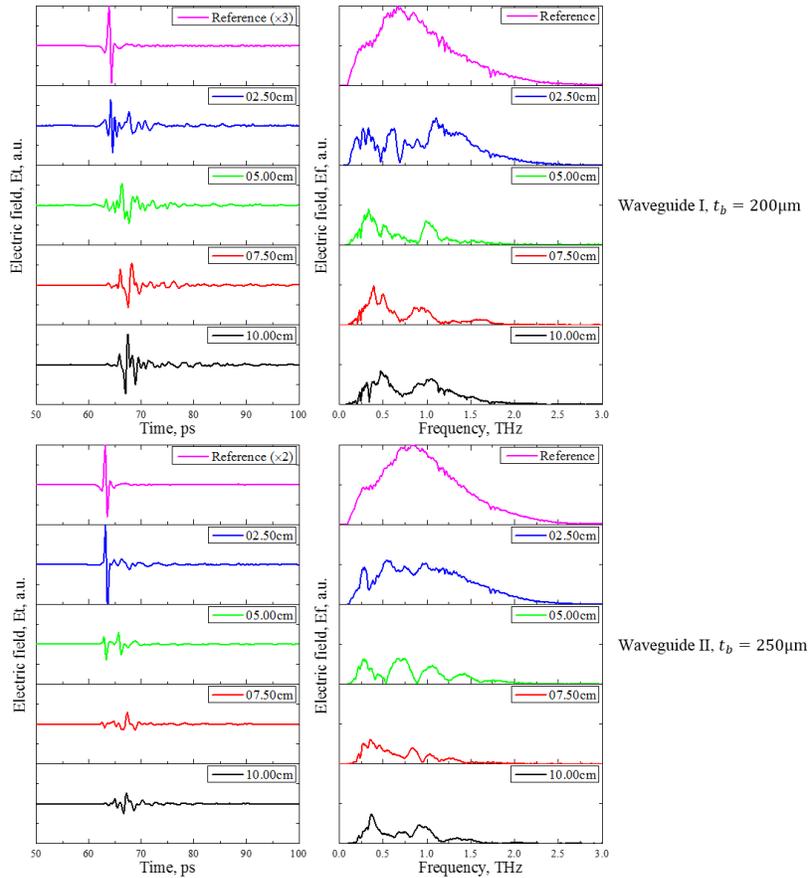

Fig. 8. Experimentally measured electric field THz traces of the two waveguide with bridge thickness of (a) 200µm and (b) 250µm. Colors of the lines refer to waveguides of different length.

To calculate the bandgap width $\Delta\omega$, we applied the second moment method detailed in [24] with the full band gap width defined as

$$\Delta\omega^2 = 4\frac{\int(\omega-\omega_c)^2 T(\omega)^2 d\omega}{\int T(\omega)^2 d\omega} \quad (12)$$

where $\omega_c$ is the bandgap central frequency and $T(\omega)$ is the field transmission. For the waveguide I with the bridge thickness of 200µm, there are four PBGs centered at frequencies of 0.17THz, 0.22THz and 0.29THz and 0.38THz characterised by enhanced transmission. The spectral width of these PBGs are 18GHz, 22GHz, 44GHz and 49GHz. Meanwhile, in the case of the waveguide II with the bridge thickness of ~250µm, four band gaps are centered at 0.14THz, 0.17THz, 0.23THz and 0.29THz, respectively. The estimated spectral widths of these band gaps are 7GHz, 25GHz, 15GHz, and 45GHz. The full width of these measured transmission bands are in good agreements with the numerical simulations presented in Fig. 5.

As the light is guided in the hollow core, transmission losses of the fabricated waveguides can be expected to be much less than that of the reflector material. From Fig. 9 we can deduce the waveguide transmission loss in various bandgap regions by comparing transmission through waveguides of different lengths. In what follows we use waveguides of 10cm and 7.5cm in our estimations. For instance, at 0.23THz, the absorption loss of the waveguide I ($t_b = 200\mu m$) is estimated to be ~0.10cm$^{-1}$, while that of the waveguide II ($t_b = 250\mu m$) is ~0.06cm$^{-1}$. As expected, the propagation losses of the two fabricated waveguides are much smaller than the corresponding bulk absorption loss of the reflector material at the corresponding frequency, which is ~0.55cm$^{-1}$ at 0.23THz according to Eq. [11].

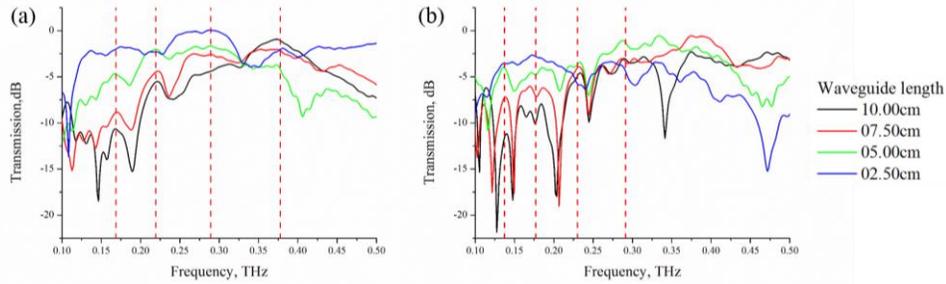

Fig. 9. The transmission spectrum of tested waveguides with bridge thickness of (a) 200µm and (b) 250µm. Red dashed lines refer to the central positions of these bandgaps

### 4. Discussion

Finally, we would like to compare our hollow-core waveguide that features a hyperuniform cladding with other hollow-core THz waveguides. In Table 1, we present bandgap width and transmission losses of several all dielectric hollow-core photonic bandgap and antiresonant waveguides. The bandgap width, is defined as the ratio of the full bandgap width to the bandgap central frequency ($\Delta\omega/\omega_0$) presented in this work (20% at 0.4THz) is comparable to the largest bandgaps reported for the photonic bandgap holey or Bragg fibers [4-14]. At the same time, it is not as wide as spectral width of regions of high transmission featured by the antiresonant waveguides [25-28]. This is to be expected as in the case of antiresonant waveguides the concept of "bandgap" is not strictly defined, and their transmission spectra do not manifest an abrupt transition from the guide to non-guided regimes when crossing the bandgap boundaries. We also believe that hyperuniform fibers can be further optimised to result in larger bandgaps via exploration of other point patterns. Thus, further work is still necessary to find out the maximal bandgap size possible with hyperuniform disordered reflectors.

We would also like to comment on the 3D stereolithography technique that has been used in this paper. Overall, this is a mature technology that has been widely used in the fabrication of microelectronic and optical devices [29-31]. Currently, ~200μm lateral resolution is standard, while some commercial system also offer sub-50μm lateral resolution. The biggest limitations of this technology are a limited material combination (resins used in fabrication), as well as limited build volume (~10-20cm linear dimension). Compared with the traditional fiber drawing fabrication method even when supplemented with stacking method [28], drilling method [32], and extrusion moulding method [33] for preform fabrication, the 3D stereolithography technology still enables fabrication of waveguides with significantly more complex transverse profile. At the same time, fiber drawing does not suffer from "resolution" issues that are present in 3D printing, thus enabling fabrication of long (meters), very smooth, submicro-thin THz structures.

| Fiber type | Reflector structure | Reflector material | Core diameter | Bandgap | Loss (cm$^{-1}$) | Ref |
|---|---|---|---|---|---|---|
| Hyperuniform fiber | Hyperuniform | Resin/air | ~5mm | 20% at 0.4THz | ~0.3 at 0.22THz | - |
| Bragg fiber | All solid | PVDF/PC | ~1mm | - | <0.02 at (1-3)THz | [4] |
| | Doped polymer | PE/PE with 80% wt. TiO$_2$ | 6.63mm | ~13% at 0.68THz | <0.042 at 0.69THz | [5] |
| | Randomly porous layer | PE/air | 6.73mm | ~12% at 0.82THz | <0.028 @ 0.82THz | [7] |
| | Porous rings | PMMA/air | 2mm | - | <1.1 (1.0-1.6THz) | [8] |
| | Cob-web structure | HDPE/air | 16mm | - | 5.84× 10$^{-6}$ at 0.55THz | [9] |
| Holey fiber | Rectangular Lattice | PTFE/air | 1.12×1.87mm | ~20% at 1.66THz for d/Λ=0.96 | - | [11] |
| | Hexagonal Lattice with regular holes | HDPE/air | 292μm | ~7.5% at 1.47THz for d/Λ=0.93; ~14% at 1.66THz for d/Λ=0.96 | ~0.022 at 1.53THz for d/Λ=0.93; ~0.014 at 1.75THz for d/Λ=0.93; | [12] |
| | Hexagonal Lattice with inflated holes | Teflon/air | 840μm | ~17% at 1.80THz | <0.04cm$^{-1}$ at (1.65-1.95)THz | [11] |
| | Honeycomb | Topas/air | ~1mm | ~4% at 0.98THz | ~0.058 at 0.98THz | [13] |
| ARROW fiber | Hollow core tube | PTFE | ~8.24mm | ~41% at 1.25 THz for 0.3mm thin tube | - | [25] |
| | Kagome | PMMA/air | 1.6mm 2.2mm | ~28% at 0.87THz ~45% at 0.77THz | <0.1 at (0.75-1)THz <0.06 at (0.65-1THz) | [26] |
| | Tube (Single ring) | PMMA/are | 1.62mm | ~23% at 0.85THz | ~0.04 at 0.83THz | [27] |
| | Tube (Several rings) | PE/air | 5.5mm | ~8% at 0.49THz | - | [28] |

Table 1. Comparison of bandgap widths and losses of the hollow core fibers with different reflector types.

## 5. Conclusion

Hollow core waveguides featuring a hyperuniform disordered reflector in the cladding are proposed for applications in the terahertz frequency range. The reflector comprises randomly positioned dielectric cylinders that are connected with thin dielectric bridges. Center positions of the cylinders follow a hyperuniform disordered point pattern which was produced using the algorithm discussed in [14] with the hyperuniformity $\chi$ is equal to 0.5. The proposed reflector structure was further optimized to maximize the photonic bandgap in the vicinity of 0.4THz, and the optimal bridge thickness and the cylinder diameter were found to be 35µm and 113µm, respectively. The resulting PBG features a full width of 80GHz (~20%).

Based on the numerically optimised waveguide structure, two hollow core waveguides with different bridge thicknesses were fabricated using a commercial 3D stereolithography printer. Due to resolution limitation of the printer, the resultant waveguides featured much wider bridge thickness (200µm and 250µm), while having the same overall structure of the reflector. We then theoretically investigated modal properties of the fabricated waveguides using a finite element method. Because of the suboptimal bridge thicknesses used in these waveguide, they exhibit smaller PBGs when compared to those of the optimal structure. Finally, we performed optical characterization of the two fabricated waveguides using a modified THz TDS system. For the fabricated waveguide with bridge thickness of 200µm, transmission band gaps are allocated at 0.17THz, 0.22THz and 0.29THz and 0.38THz, and the corresponding spectral widths are 18GHz, 22GHz, 44GHz and 49GHz, respectively, resulting a maximum band gap of ~15.1% at 0.29THz. When the bridge thickness is increased to 250µm, central frequencies of these bands are shifted to 0.14THz, 0.17THz, 0.23THz and 0.29THz, respectively, and the corresponding spectral widths are reduced to 7GHz, 25GHz, 15GHz, and 45GHz. The maximum bandgap of this waveguide is measured to be ~15.3% at 0.29THz. The location and the widths of the experimentally measured band gaps are in good agreement with the theoretical prediction. Moreover, due to hollow core guidance, transmission losses (within then bandgaps) of the fabricated waveguides are significantly smaller than the bulk absorption loss of the reflector material.